Review article

Pierre Türschmann, Hanna Le Jeannic, Signe F. Simonsen, Harald R. Haakh[a], Stephan Götzinger, Vahid Sandoghdar, Peter Lodahl and Nir Rotenberg*

# Coherent nonlinear optics of quantum emitters in nanophotonic waveguides

**Abstract:** Coherent quantum optics, where the phase of a photon is not scrambled as it interacts with an emitter, lies at the heart of many quantum optical effects and emerging technologies. Solid-state emitters coupled to nanophotonic waveguides are a promising platform for quantum devices, as this element can be integrated into complex photonic chips. Yet, preserving the full coherence properties of the coupled emitter-waveguide system is challenging because of the complex and dynamic electromagnetic landscape found in the solid state. Here, we review progress toward coherent light-matter interactions with solid-state quantum emitters coupled to nanophotonic waveguides. We first lay down the theoretical foundation for coherent and nonlinear light-matter interactions of a two-level system in a quasi-one-dimensional system, and then benchmark experimental realizations. We discuss higher order nonlinearities that arise as a result of the addition of photons of different frequencies, more complex energy level schemes of the emitters, and the coupling of multiple emitters via a shared photonic mode. Throughout, we highlight protocols for applications and novel effects that are based on these coherent interactions, the steps taken toward their realization, and the challenges that remain to be overcome.

## 1 Introduction

Quantum optical research no longer focuses solely on fundamental demonstrations of the quantum nature of atoms, photons, or their interactions, but rather integrates these constituents into increasingly complex systems. These provide an ever-growing view of the rich realm of many-body quantum physics [1] and bring us closer to functional quantum technologies such as quantum networks [2–4] and, ultimately, quantum computers [5, 6].

One basic element that has the potential to fulfill many of the functionalities required in complex, active quantum architectures is a quantum emitter coupled to a photonic waveguide, as sketched in Figure 1. Nanophotonic waveguides confine and guide light and reshape the emission from dipole sources to match their fundamental mode [7–9], and can be engineered to strongly suppress emission into free space [10], resulting in efficient emitter-photon coupling [11]. An emitter efficiently coupled to a waveguide can therefore act as a source of high-quality single photons [12], for example, for quantum information processing with linear optical systems [13, 14]. The strong confinement of light in waveguides also leads to the presence of a large longitudinal component of the electric field, and the interaction of this vector field with circular dipoles can result in unidirectional emission [15–19].

[a]**Current address:** Federal Ministry of Education and Research, D-53170 Bonn, Germany
**\*Corresponding author: Nir Rotenberg,** Niels Bohr Institute and Center for Hybrid Quantum Networks, University of Copenhagen, Blegdamsvej 17, DK-2100 Copenhagen, Denmark,
e-mail: nir.rotenberg@nbi.ku.dk. https://orcid.org/0000-0002-8172-7222
**Pierre Türschmann:** Linnowave GmbH, Henkestr. 91, D-91052 Erlangen, Germany; and Max Planck Institute for the Science of Light, Staudtstr. 2, D-91058 Erlangen, Germany
**Hanna Le Jeannic, Signe F. Simonsen and Peter Lodahl:** Niels Bohr Institute and Center for Hybrid Quantum Networks, University of Copenhagen, Blegdamsvej 17, DK-2100 Copenhagen, Denmark
**Harald R. Haakh:** Max Planck Institute for the Science of Light, Staudtstr. 2, D-91058 Erlangen, Germany
**Stephan Götzinger:** Max Planck Institute for the Science of Light, Staudtstr. 2, D-91058 Erlangen, Germany; Department of Physics, Friedrich Alexander University Erlangen-Nürnberg (FAU), D-91058 Erlangen, Germany; and Graduate School in Advanced Optical Technologies (SAOT), FAU, D-91052 Erlangen, Germany
**Vahid Sandoghdar:** Max Planck Institute for the Science of Light, Staudtstr. 2, D-91058 Erlangen, Germany; and Department of Physics, Friedrich Alexander University Erlangen-Nürnberg (FAU), D-91058 Erlangen, Germany

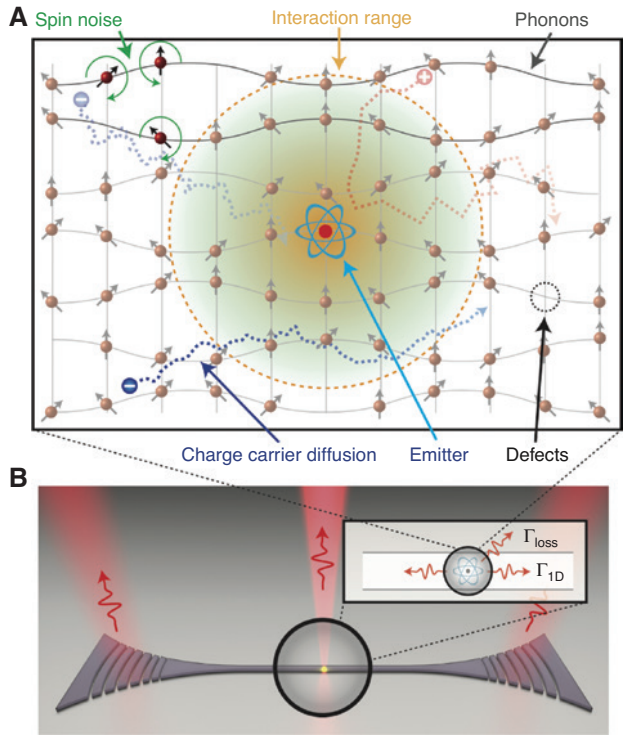

**Figure 1:** The basic system considered in this review: a quantum emitter coupled to a quasi-one-dimensional photonic waveguide. The coherence of this quantum system is degraded by the interaction of the emitter with the solid-state environment, as discussed in the text. (A) Here, sources of noise such as charge, spin, phonons, and nearby defects are schematically depicted. (B) Emission, in this system, occurs either into the guided modes, with a rate $\Gamma_{1D}$, or is lost into other modes with a rate $\Gamma_{loss}$.

These directional light-matter interactions enable the creation of chiral quantum optical elements such as optical isolators, single-photon routers, quantum logic gates, and even networks [20, 21].

Spurred by this potential, researchers have, in the past few years, worked to couple a variety of quantum emitters to waveguides, including single organic molecules [22–25], a variety of color centers in diamond [26–29], atoms [30–33], quantum dots (QDs) [9], and superconducting qubits [34]. These systems, with the exception of superconducting qubits and free-space Rydberg-atom quantum optics that are covered by the latter review, have, over the last decade, been brought to the nanoscale.

The challenge to interfacing solid-state emitters with nanophotonic waveguides is in keeping the ensuing light-matter interactions coherent. As sketched in Figure 1A, there are many possible sources of noise in solid-state systems, from ballistic or trapped charges near the emitter, to spin noise in the surrounding nuclear bath, to the coupling to phonons or to vibration in the vicinity of the emitter [35–46]. The methods implemented to overcome these processes depend on the type of quantum emitter and include, for example, the careful crystallization of the host matrix of single organic molecules [22, 45], embedding epitaxially grown QDs in a diode to shield them from electronic fluctuations [47, 48], using Purcell enhancement to overcome dephasing [26, 49, 50], and searching for better shielded defects within diamond [51].

Here, we review the current state of the art in coherent quantum optics in nanophotonic waveguides, focusing on quantum optical nonlinearities. We begin with the theoretical background that describes coherent quantum optics of two-level systems in one dimension, highlighting the important parameters and figures of merits associated with these interactions, and the corresponding experimental demonstrations. We then review higher order quantum optical nonlinearities, touching on effects that require multiple photons and extending beyond two-level systems, before concluding with a discussion of coherent quantum optics in multi-emitter systems.

## 2 Nonlinear response of a two-level system coupled to a waveguide

### 2.1 Transmission and reflection

Of the many theories developed to describe quantum light-matter interactions [52], the Green's function formalism is particularly well suited to describe the coherent interaction of guided photons with two-level systems (TLSs) [53]. This formalism allows for a full quantum treatment of dispersive and absorbing open systems and therefore spans both plasmonic and dielectric waveguides and, importantly, can be extended to multi-emitter systems (Section 4). In this section, we briefly outline this formalism, focusing on measurable signatures of coherent light-matter interactions with TLSs embedded in one-dimensional waveguides.

The Hamiltonian describing the interaction of a TLS with light in a single photonic mode, in a reference frame rotating at the angular frequency of the light field $\omega_P$, is

$$\hat{\mathcal{H}} = -\hbar\Delta_P \hat{\sigma}_{eg}\hat{\sigma}_{ge} + \hbar\omega_P \hat{\mathbf{f}}^\dagger(\mathbf{r})\hat{\mathbf{f}}(\mathbf{r}) - \hat{\mathbf{d}}\cdot\hat{\mathbf{E}}(\mathbf{r}), \quad (1)$$

where $\Delta_P = \omega_P - \omega_A$. The first term describes the TLS, whose transition energy is $\hbar\omega_A$ and whose coherences between the $i$ and $j$ levels are given by $\hat{\sigma}_{ij} = |i\rangle\langle j|$. Likewise, the second term relates to the excitation of the light field, here taken to have energy $\hbar\omega_p$, described by the creation and annihilation bosonic operators $\hat{f}$ and $\hat{f}^\dagger$.

The last term of Eq. (1) describes the light-matter interaction, which is mediated by the transition dipole of the emitter. Here, the dipole operator can be written in terms of the dipole matrix elements $\mathbf{d}=\langle g|\hat{\mathbf{d}}|e\rangle$ as $\hat{\mathbf{d}}=\mathbf{d}^*\hat{\sigma}_{eg}+\mathbf{d}\hat{\sigma}_{ge}$ and $\hat{\mathbf{E}}(\mathbf{r})$ is the electric field operator, to which we will return shortly.

First, however, it is instructive to consider solely the state of the emitter, using the reduced density matrix for the TLS $\hat{\rho}$, which is related to the expectation values of the atomic operators through $\langle\hat{\sigma}_{ij}\rangle=\rho_{ji}$. The emitter density matrix operator evolves according to [54]

$$\frac{d\hat{\rho}}{dt}=-\frac{i}{\hbar}[\hat{\mathcal{H}},\hat{\rho}]+\mathcal{L}[\hat{\rho}], \qquad (2)$$

where the Lindblad operator for the single emitter system [55, 56]

$$\mathcal{L}[\hat{\rho}]=\sum_{ij}\frac{\Gamma_{ij}}{2}(2\hat{\sigma}_{ji}\hat{\rho}\hat{\sigma}_{ij}-\hat{\sigma}_{ii}\hat{\rho}-\hat{\rho}\hat{\sigma}_{ii}) \qquad (3)$$

accounts for both the decay and decoherence of the emitter. For the emitter-waveguide system, this operator has three nonzero terms: $\Gamma_{eg}\equiv\Gamma$, which is the spontaneous emission rate associated with the transition from $|e\rangle$ to $|g\rangle$, and $\Gamma_{ee}=\Gamma_{gg}\equiv\Gamma_{deph}$, which is the pure dephasing of the system through which it decoheres without undergoing a transition.

Equation (2) can be solved using the rotating wave approximation and by assuming Markovian dephasing processes to yield the steady-state $\left(\frac{d}{dt}\hat{\rho}=0\right)$ elements of the reduced density matrix

$$\rho_{ee}=\frac{2\Gamma_2\Omega_P^2}{\Gamma(\Gamma_2^2+\Delta_P^2+4(\Gamma_2/\Gamma)\Omega_P^2)}, \qquad (4)$$

$$\rho_{ge}=-\frac{\Omega_P(i\Gamma_2+\Delta_P)}{\Gamma_2^2+\Delta_P^2+4(\Gamma_2/\Gamma)\Omega_P^2}, \qquad (5)$$

where $\Gamma_2=\Gamma/2+\Gamma_{deph}$ and we have defined the Rabi frequency to be $\Omega_P=\mathbf{d}\cdot\mathbf{E}/\hbar$ for the driving field amplitude at the position of the emitter $\mathbf{E}=\langle\hat{\mathbf{E}}\rangle$. Note that the other two elements of the reduced density matrix are simply $\rho_{gg}=1-\rho_{ee}$ and $\rho_{eg}=\rho_{ge}^*$. Interestingly, Eq. (4) describes the spontaneous emission of the emitter, showing both the amplitude saturation and linewidth broadening as a function of $\Omega_P$.

To understand how the emitter interacts with photons propagating through the waveguide, we return to Eq. (1), now writing the electric field operator as $\hat{\mathbf{E}}(\mathbf{r},\omega)=\hat{\mathbf{E}}^+(\mathbf{r},\omega)+\hat{\mathbf{E}}^-(\mathbf{r},\omega)$, where, in general [57]

$$\hat{\mathbf{E}}^+(\mathbf{r},\omega)=i\mu_0\sqrt{\frac{\hbar}{\pi\varepsilon_0}}\frac{\omega^2}{c^2}\int d\mathbf{r}'\sqrt{\varepsilon_I(\mathbf{r}',\omega)}\,\mathbf{G}(\mathbf{r},\mathbf{r}',\omega)\cdot\hat{\mathbf{f}}(\mathbf{r}',\omega), \qquad (6)$$

and the negative frequency component of the field operator, $\hat{\mathbf{E}}^-(\mathbf{r},\omega)$, is the Hermitian conjugate of Eq. (6). Here, we explicitly note the frequency dependence of the different quantities (which, henceforth, will be removed for clarity and understood to be evaluated at $\omega_P$). The presence of the imaginary component of the dielectric function $\varepsilon_I$ alongside $\hat{\mathbf{f}}$, as required by the fluctuation dissipation theorem [58], ensures that this formalism is valid for dispersive and absorbing systems. Intuitively, this equation states that the field at any position $\mathbf{r}$ is comprised of photons emitted at all positions $\mathbf{r}'$, which then propagate back to $\mathbf{r}$. This process is described by the Green's tensor $\mathbf{G}(\mathbf{r},\mathbf{r}',\omega)$, meaning that photons propagate through the system in the same manner as classical electromagnetic waves.

The dipole-projected Green's function for a one-dimensional waveguide, where light propagates in the $z$ direction, is [59, 60]

$$\begin{aligned}\mathbf{g}(\mathbf{r}_i,\mathbf{r}_j)&=\frac{\mu_0\omega_A^2}{\hbar}\mathbf{d}^*(\mathbf{r}_i)\cdot\mathbf{G}(\mathbf{r}_i,\mathbf{r}_j)\cdot\mathbf{d}(\mathbf{r}_j),\\ &\approx i\frac{\beta\Gamma}{2}e^{ik_P|z_i-z_j|},\end{aligned} \qquad (7)$$

where $k_P$ is the wavenumber of the photonic mode. Here, we define the emitter-waveguide coupling efficiency $\beta\equiv\Gamma_{1D}/\Gamma$ as the emission rate into the guided mode $\Gamma_{1D}$ normalized by the total emission rate $\Gamma=\Gamma_{1D}+\Gamma_{loss}$ (c.f. inset to Figure 1B). Following some algebra, this equation for $\mathbf{g}(\mathbf{r}_i,\mathbf{r}_j)$ allows us to rewrite Eq. (6) in terms of the incident field $\hat{\mathbf{E}}_P^+$ and scattered field

$$\hat{\mathbf{E}}^+(\mathbf{r})=\hat{\mathbf{E}}_P^+(\mathbf{r})+i\frac{\beta\Gamma}{2\Omega_P}\hat{\mathbf{E}}_P^+(\mathbf{r})\hat{\sigma}_{ge}. \qquad (8)$$

Equations (4), (5), and (8) allow us to quantify the light-matter interaction of the TLS with guided photons. The transmission through the waveguide, for example, is $T=\langle\hat{\mathbf{E}}^-\hat{\mathbf{E}}^+\rangle/\langle\hat{\mathbf{E}}_P^-\hat{\mathbf{E}}_P^+\rangle$, which can be written using Eq. (8) as

$$T=1+\left(\frac{\beta\Gamma}{2\Omega_P}\right)^2\rho_{ee}+i\frac{\beta\Gamma}{2\Omega_P}(\rho_{eg}-\rho_{ge}). \qquad (9)$$

The first term of this equation, which depends on the population of the excited state of the emitter, is typically thought to govern the incoherent interactions, while the second term depends on the atomic coherences and is therefore viewed as the source of the coherent interactions. This view is particularly attractive since $\langle \hat{\sigma}_{ee} \rangle$ dominates over $\langle \hat{\sigma}_{eg} \rangle$ at high energies [c.f. Eqs. (4) and (5)]. In light of these equations and the prefactors of Eq. (9), however, it is clear that both terms have the same power dependence, and it is not so simple to separate the coherent and incoherent contributions to the transmitted field.

Such a separation is, however, relatively straightforward in the low-power and no-detuning limit, where we can use Eqs. (4) and (5) to rewrite the transmission as

$$T \approx 1 - \frac{\Gamma}{\Gamma_2}\beta + \frac{\Gamma}{2\Gamma_2}\beta^2, \quad (10)$$

where it is clear that, in the limit of no pure dephasing, all terms contribute. If we then define the fraction of coherent interactions to be $\beta_{co} \equiv \Gamma_{1D}/2\Gamma_2$, we can rewrite the low-power transmittance as

$$T \approx (1-\beta_{co})^2 + \beta_{co}(\beta - \beta_{co}), \quad (11)$$

where the first and second terms are the coherent and incoherent contributions.

More generally, the transmission expressed in terms of the system parameters is

$$T = 1 - \frac{\beta\Gamma\Gamma_2(2-\beta)}{2(\Gamma_2^2 + \Delta_P^2 + 4(\Gamma_2/\Gamma)\Omega_P^2)}. \quad (12)$$

Similarly, the reflection from the TLS is

$$R = \frac{\beta^2\Gamma\Gamma_2}{2(\Gamma_2^2 + \Delta_P^2 + 4(\Gamma_2/\Gamma)\Omega_P^2)}, \quad (13)$$

and, in the absence of absorption, the losses due to the scattering of light out of the waveguide mode is simply $S = 1 - T - R$. Note that these equations have been derived for excitation by a weak, continuous wave beam, but they also hold for spectrally narrow single photons (in the limit of $\Omega_P \ll 1$) [61].

A clear signature of the coherent interaction between a TLS and an emitter can be found in the transmission (and reflection) signals [62]. For a perfect system ($\beta \to 1$, $\Gamma_{deph} \to 0$) in the low-power limit ($\Omega_P \to 0$), all single photons are reflected, leading to a perfect extinction of the transmission $\Delta T = T(\Delta_P = \pm\infty) - T(\Delta_P = 0) = 1$. As is evident from Eq. (12), this extinction depends nonlinearly on the power, coupling efficiency, and dephasing rate of the emitter. Perfect extinction is not possible with tightly focused plane waves, where a theoretical maximum of $\Delta T = 0.85$ has been calculated [63]. Rather, perfect extinction in free space requires perfectly matching the dipole mode with the excitation beam, a notoriously difficult proposition that motivates the importance of nanophotonic platforms. Waveguides, as we noted earlier, both reshape the radiation pattern of dipoles to match their fundamental modes [7, 8] and are non-diffracting, meaning that these structures are particularly well suited to ideally interface with quantum emitters.

Coherent extinction has recently been observed in a variety of systems, as summarized in Figure 2 and Table 1. Multiple organic molecules have been coupled coherently to a single nanoguide, each of which has a near-lifetime-limited transition that can be addressed individually through spectral-spatial selection [22] (Figure 2A). Extinction up to $\Delta T \approx 0.09$ has been reported for this system [23]. Likewise, $\Delta T \approx 0.18$ has been measured for Ge vacancies, deterministically implanted in a diamond waveguide (Figure 2B) [28], and $\Delta T \approx 0.67$ for InAs QDs embedded in a gated nanobeam waveguide [64] (Figure 2C). In all cases, the limiting factor has been the $\beta$ of the systems [c.f. Eq. (12)]. For nanoguides, such as those used in the aforementioned experiments, the maximum achievable $\beta$ depends on the confinement of the guided mode, which, in turn, is a function of the refractive index ratio between the waveguiding medium and its surrounding, and the geometrical size of the waveguide. These dependencies are plotted in Figure 2D, where $\beta$ is semianalytically calculated for nanoguides with circular cross-sections, showing the largest possible $\beta$ for the different quantum emitters.

Alternatively, $\beta$ could be increased through the use of photonic resonators, or highly structured systems such as photonic crystals. The generalization of Eqs. (12) and (13) to account for cavity Purcell enhancement is relatively straightforward and leads to the observation of Fano-like lineshapes in the transmission, as the phase of the photons that do not interact with the emitter is now dependent on their spectral detuning from the resonance [75]. Coherent and deterministic light-matter interactions have, in fact, recently been demonstrated with a single molecule [74], defect centers [68], and QDs [72] in microcavities, where a $\Delta T \approx 0.99$ has been observed, albeit at the cost of operational bandwidth. Here, the emission into the photonic mode was Purcell enhanced, increasing the emission rate in the desired channel relative to other radiative channels and the dephasing rate. The nonlinear dependence of $\Delta T$ has been observed for both molecules [23] and QDs [64, 69, 70], where a critical flux of $\approx 1$ photon per lifetime was found to saturate the emitter, as shown in Figure 3A, since it can only scatter a single photon at a time. In a complementary fashion, the phase of a scattered photon can be tuned, as the presence of a TLS fundamentally changes the response of a nanophotonic system, as shown in Figure 3B [67]. Altogether, these results hint at the power of coherent quantum optics in integrated photonic systems.

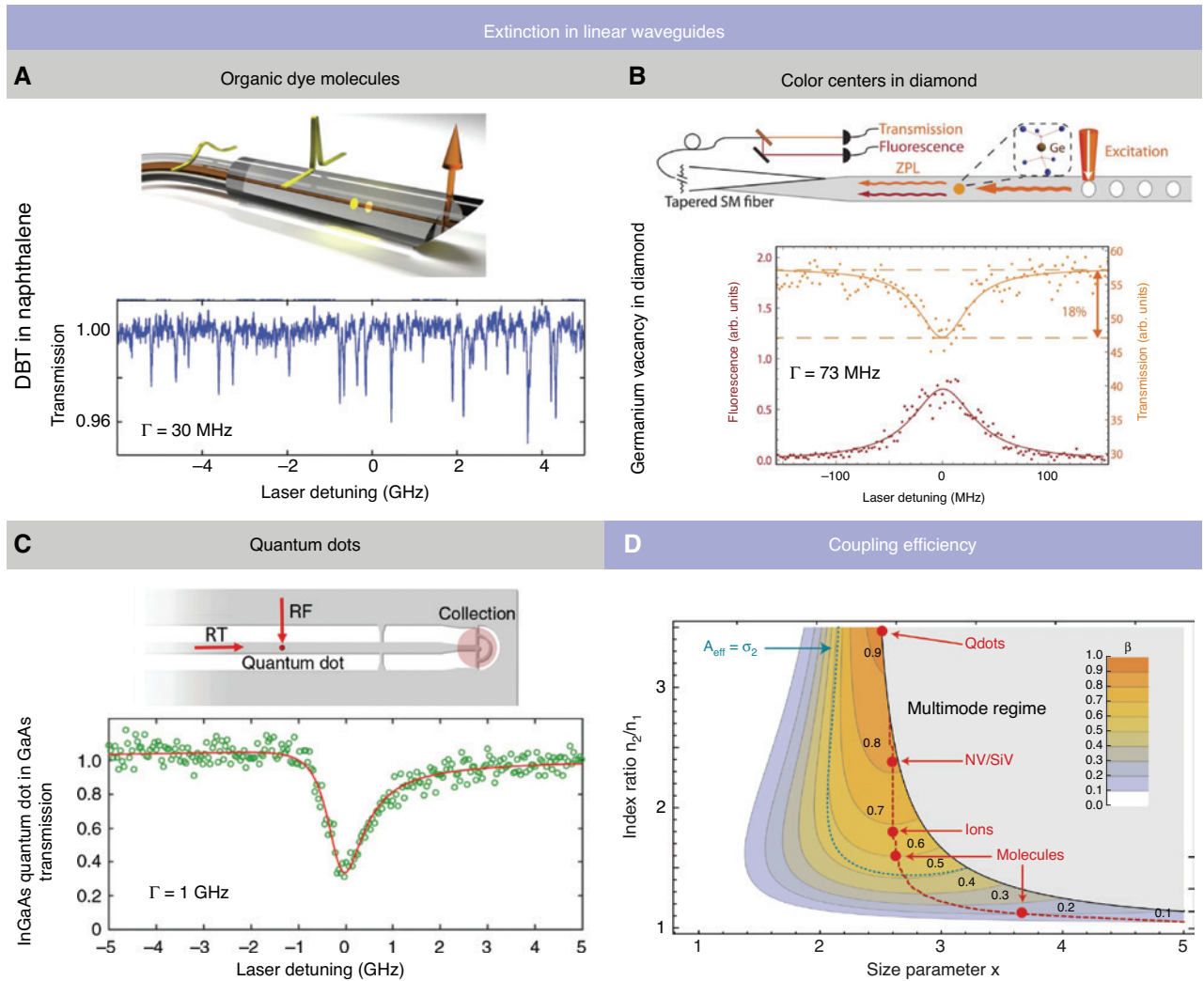

**Figure 2:** State-of-the-art coherent extinction with a variety of solid-state quantum emitters in differing nanoguide geometries, including (A) single organic molecules, (B) Ge defects in diamond, and (C) InAs quantum dots. Adapted from [22], [28], and [64], respectively. (D) Maximum achievable coupling coefficient $\beta$ for cylindrical nanoguides of different core sizes and the refractive index contrast surrounding the core. The core size is parameterized by the unitless variable $x = k_2 r$, where $k_2$ is the wavenumber of light in the bulk material of the core and $r$ is the radius of the nanoguide. Achievable $\beta$-factors for the different solid-state quantum emitters are marked in red symbols (dashed red line denotes the optimal achievable $\beta$ for a given refractive index contrast), as is the geometry for which the effective mode area is equal to the scattering cross-section of an emitter (dashed blue line).

## 2.2 Nonlinearity and photon statistics

A more profound signature of the coherent nonlinearity of the TLS can be found in the ensuing photon statistics, which explicitly demonstrates the different response of the TLS to single and multiple incident photons. This is seen in the normalized second-order correlation function

$$g^{(2)}(\tau) = \frac{G^{(2)}(t, \tau)}{[G^{(1)}(t)]^2}, \quad (14)$$

where $G^{(1)}(t) = \langle \hat{\mathbf{E}}^-(t)\hat{\mathbf{E}}^+(t)\rangle$, and $G^{(2)}(t, \tau) = \langle \hat{\mathbf{E}}^-(t)\hat{\mathbf{E}}^-(t+\tau)\hat{\mathbf{E}}^+(t+\tau)\hat{\mathbf{E}}^+(t)\rangle$. The field operator used in these correlation functions is given in Eq. (8), but here $\hat{\sigma}_{ge}(t)$ evolves in time. For continuous wave excitation, the first-order correlation function can be simply derived by noting that in the steady-state $\rho_{ij} = \langle \hat{\sigma}_{ji}(0)\rangle$ and by using Eqs. (4) and (5).

Table 1: State-of-the-art coherent light-matter interactions with solid-state quantum emitters coupled to nanophotonic waveguides.

| System | Extinction | Linewidth[a] | $g^{(2)}(0)$[b] | Details | Reference |
|---|---|---|---|---|---|
| Atoms coupled to photonic waveguides | | | | | |
| Cs-nanoguide | 0.01 | 5.8 (5.8) MHz | – | More than 2000 resonantly coupled emitters | [30, 65] |
| Cs-PhCW | 0.25 | 15 (4.6) MHz | – | Mean number of coupled atoms is 3; observation of superradiance | [33] |
| Rb-nanoguide | 0.20 | 6.1 (6.1) MHz | – | 1-6, mean number of coupled atoms; observation of superradiance | [66] |
| Rb-PhCC | – | 53 MHz | 0.12(4.1) | Nanophotonic control of photon phase | [67] |
| Defect centers in diamond | | | | | |
| GeV-PhCW | 0.18 | 73(26) MHz | <0.08(1.1) | Coherent nonlinearity at the single-photon level | [28] |
| SiV-PhCW | 0.38 | 590(90) MHz | <0.15(1.5) | Two emitters remotely entangled by Raman transitions | [26] |
| SiV-PhCC | >0.95 | 4.6 GHz | 0.23 | Two Si vacancies, near-field-coupled inside a single cavity | [68] |
| InAs QDs | | | | | |
| QD-nanobeam | 0.66 | 1.2(0.9) GHz | <0.01 | Charge-stabilized and tunable by a diode | [12, 64] |
| QD-PhCW | 0.07–0.35 | 1.1–4(0.9) GHz | <0.01(1.15) | Coherent nonlinearity at the single photon level | [69, 70] |
| QD-PhCW | 0.85 | 1.36(1.22) GHz | (6) | Charge-stabilized and tunable by a diode | [71] |
| QD-microcavity | – | 0.28(0.28) GHz | (25–80) | Charge-stabilized in a $\mu$-pillar or $\mu$-cavity diode | [72, 73] |
| Organic molecules | | | | | |
| DBT-nanoguide | 0.07–0.09 | 30(30) MHz | <0.01 | Observation of up to 5000 single molecules coupled to the waveguide | [22, 23] |
| DBT-microcavity | 0.99 | 40(40) MHz | 0.1(21) | Measured $g^{(2)}(0)$ limited by timing-resolution of the detectors | [74] |

Atoms are included for completeness.
PhCW, Photonic crystal waveguide; PhCC, photonic crystal cavity.
[a]Natural linewidth, which may be Purcell-enhanced, is given in brackets.
[b]Bunching value observed in coherent transmission experiments is given in brackets.

Similarly, the quantum regression theorem [54] allows us to express the two-time correlations found in $G^{(2)}(\tau)$ in terms of the steady-state elements of the density matrix. In this manner, and in the low-excitation limit ($\Omega_P \to 0$), we can write

$$g^{(2)}(\tau) = 1 + \frac{\Gamma^2 \Gamma_2 \beta^4 e^{-\Gamma \tau} - \Gamma \beta^2 [2\Gamma_2 + \Gamma(\beta-2)]^2 e^{-\Gamma_2 \tau}}{(\Gamma - \Gamma_2)[2\Gamma_2 + \Gamma\beta(\beta-2)]^2}. \quad (15)$$

In the limit of no pure dephasing ($\Gamma_2 \to \Gamma/2$), this equation converges to that of Chang et al. [76].

The highly nonlinear response of a TLS is encoded into the photon statistics of the transmitted light. A signature of this nonlinearity is the photon bunching observed in $g^{(2)}(0)$, which we plot in Figure 4A as a function of both $\beta$ and the relative dephasing rate $\Gamma_{\text{deph}}/\Gamma$. The strong photon bunching is observed when $\beta \to 1$ and $\Gamma_{\text{deph}} \to 0$. Here, $g^{(2)}(0) \to \infty$, as $T \to 0$ (c.f. Section 2.1), meaning that all single-photon components of the incident coherent state are reflected. Conversely, in transmission, the coherent superposition of the zero and multi-photon components results in photon bunching. This photon bunching has been observed with both QDs [69–71] and Ge vacancy centers [28] coupled to waveguides, with peak values of $g^{(2)}(0)$ ranging from 1.1 to 6. Higher values have recently been reported using a variety of emitters coupled to resonators, as summarized in Table 1. The photon statistics can be actively manipulated, for example, to alternate between bunching and antibunching, using the higher order nonlinearity of the emitter [77] or by modulating the local phase of the quantum interference [78, 79].

This multi-photon transmission is comprised of two components: one in which the two photons are uncorrelated and therefore scatter independently, and one where the photons are correlated and cannot be considered independently [63, 80–82]. Shen and Fan derived analytic formulas for both components using a scattering matrix approach, and later using the input-output formalism [83]. Subsequently, Ramos and Garcia-Ripoll proposed an experimental method to measure the single- and two-photon scattering matrices using weak coherent beams [84].

A part of the correlated component of the two-photon wavepacket whose shape does not change as a result of scattering from the TLS exists. This component acts like a "quantum soliton" and is known as a photon-photon bound state. This is much like photon-emitter bound states, which are characterized by the entanglement between the light and matter degrees of freedom [85, 86], a hallmark

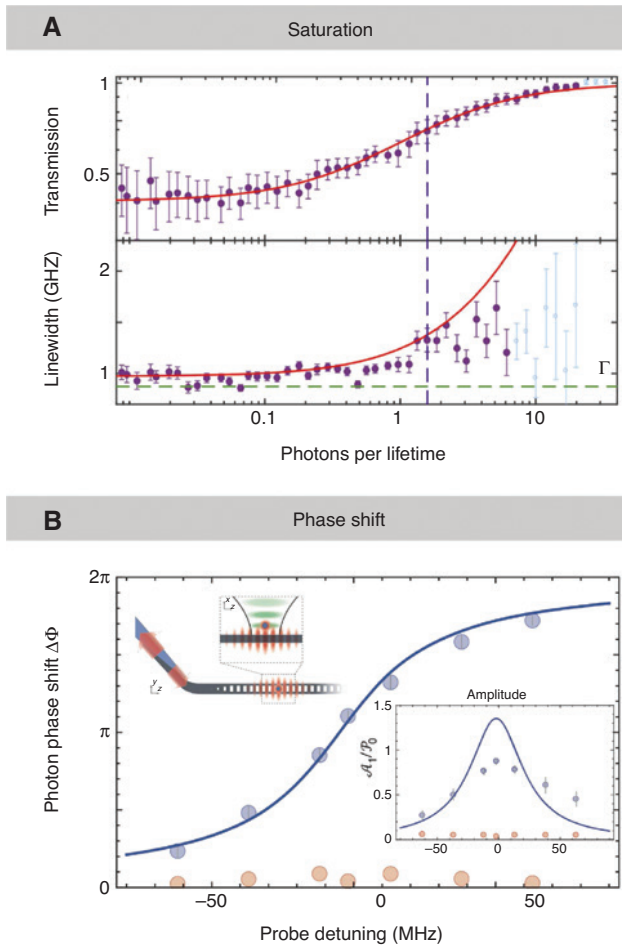

The highly nonlinear coherent scattering of guided photons from single emitters, and the ensuing strong correlations of photon-photon bound states, constitute a valuable quantum resource and have played a central role in recent proposals. The large disparity between the response of the emitter to single- and two-photon inputs can form the basis for a photon sorter, allowing for the realization of quantum nondemolition measurements [93] and for the creation of a Bell-State analyzer and controlled-sign gate [94]. This latter proposal may benefit from chiral coupling, meaning that all photons scatter in a single (forward) direction [21]. Interestingly, if this coupling could be made asymmetric but not perfectly directional, then this passive two-level nonlinearity has been predicted to act as a few-photon diode [95].

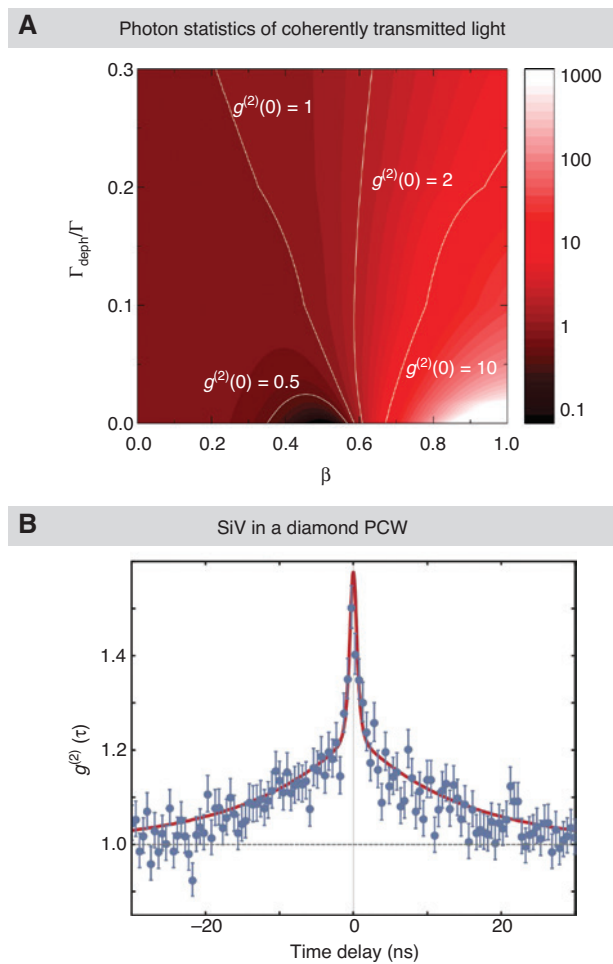

**Figure 3:** Nonlinear dependence of $\Delta T$ and the presence of a TLS fundamentally changes the response of a nanophotonic system.
(A) The coherent nonlinear response of a TLS is seen in the power-dependent extinction of the transmitted light, which vanishes as the photon flux increases beyond one photon per lifetime (top panel). As the same time, the bandwidth of the transition begins to power-broaden (bottom panel), signifying the loss of coherence in the light-matter interactions. Adapted from [64]. (B) An efficiently coupled TLS, here an atom evanescently coupled to a photonic crystal cavity (top inset), can also modulate the phase of the scattered photons. In fact, markedly different responses are seen in the presence of the emitter (blue symbols) and in its absence (yellow symbols). The bottom inset shows the corresponding normalized count rate in one arm of an interferometer, relative to the expected cavity response (blue curve). Adapted from [67].

of polaritonics [87]. The work on photonic bound states was generalized to higher number multi-photon bound states [88] and to their spectral and temporal signatures in structured photonic environments [89]. Experimental signatures of the correlated components of two [90] and three [91, 92] photon scattering events were recently reported with Rydberg atoms, but an unambiguous observation of a photonic bound state is, to date, missing.

**Figure 4:** Two-photon correlations for a weak coherent-state scattering from a TLS embedded in a waveguide.
(A) $g^{(2)}(0)$ as a function of both $\beta$ and $\Gamma_{deph}$ [Eq. (15)], showing a strong photon bunching for well-coupled emitters with small pure dephasing. (B) Using an integrated nanophotonic waveguide cavity coupled to a Si vacancy in diamond, researchers were able to observe a $g^{(2)}(0) \approx 1.5$. Adapted from [26].

# 3 Nonlinearities of multilevel systems

In analogy to classical optics, there are a host of quantum optical nonlinearities beyond the interaction of photons with a TLS described above. These higher order quantum nonlinearities result from (i) the presence of multiple photons of different frequencies or (ii) a richer energy level structure of the emitter, and cannot be described by a perturbative susceptibility tensor like classical nonlinearities. In this section we describe these effects and highlight recent efforts to observe them in nanophotonic waveguides.

## 3.1 Dressed two-level systems

TLSs can coherently mediate the interaction between two different photons that, unlike the scenario described in Section 2, can be frequency-detuned. The resulting multi-photon nonlinear effects were first explored by Wu, Ezekiel, Duckloy, and Mollow in 1977, who measured the transmission of a weak probe beam through a large ensemble of sodium atoms in the presence of a strong excitation laser [96]. Because of the weak light-matter interactions in these initial experiments, they were conducted using large ensembles of atoms and with intense control beams. In fact, 30 years would pass before technological progress enabled the observation of these nonlinearities at the single-photon single-emitter level, as we discuss below, when spectra such as those shown in the right panel of Figure 5A were recorded using a single molecule [23, 97].

The physics underlying these spectra can be understood in terms of the three available transitions of the dressed state picture [98], where the bare states of the emitter hybridize with the manifold of light states, as shown in the left panel of Figure 5A. First, the emitter resonance, which is observed as an extinction of the transmission as in Section 2.1, is AC-Stark shifted by the presence of pump photons (red arrows). Second, when the pump and signal are only slightly detuned, the emitter can mediate the transfer of photons between the two beams, which appears as a kink in the transmission spectra (corresponding to the green transition). Finally, a stimulated process that requires two pump photons can coherently amplify the signal beam without the need for population inversion, resulting in the bump of the transmission signal (corresponding to the blue transition). A theoretical model, based on the optical Bloch equations, accurately reproduces these complex spectra [99, 100].

Initially, observing these multicolor nonlinear effects with single emitters proved challenging, because of the low probability of each photon interacting with an emitter in bulk. These constraints were first overcome using a combination of sensitive lock-in techniques and ultra-strong pump fields, to observe the signatures of these nonlinearities at the $10^{-5}$ level, first in the absorption [101] and then in the transmission [102] of a single QD in a bulk medium. Maser et al. improved this signal by three orders of magnitude by focusing tightly on a single organic molecule with a transform-limited transition, embedded in a thin organic matrix. Using this same platform, researchers were then able to observe these multicolor nonlinearities mediated by a single organic molecule coupled to a waveguide on a photonic chip [23], as shown in Figure 5B. Here, a clear nonlinear dependence of both the extinction and coherent amplification signals on the detuned pump photons is seen. Note, however, that even in this most recent experiment the extinction peaks at $\Delta T \approx 0.1$ because of a weak molecule-waveguide coupling of only $\beta = 0.08$. Regardless, this increased sensitivity both drastically reduced the amount of pump photons needed to observe these effects and allowed for the observation of additional nonlinear effects such as the four-wave mixing shown in Figure 5C [97]. Additional nonlinear frequency conversion processes have been predicted, but not yet observed, for strong driving fields and sufficiently strong coupling [103].

Experimentally, the current challenge is to reach the regime where these multicolor nonlinearities can be deterministically observed, and perhaps exploited to control light-matter interactions at the single-photon level. This requires that the light-matter coupling efficiency approaches unity while maintaining a fully coherent interaction (i.e. $\Gamma_{deph} = 0$), further motivating the use of structured nanophotonic waveguides and resonators.

## 3.2 Three- and four-level emitters

Although we often approximate quantum emitters as TLSs, they may actually have a much richer energy level structure that gives rise to new coherent and nonlinear quantum optical effects. In recent years, there have been many theoretical studies on the use of waveguides to enhance and exploit these effects, laying out the framework for future experiments.

A prototypical example of such a coherent, multi-level quantum optical nonlinearity is electromagnetically induced transparency (EIT). EIT occurs when, in the presence of a control field, destructive quantum interference between two transitions of a three-level system (c.f. Figure 6A) prevents absorption of a weak signal field [104]. The coherent control of quantum absorption enables, for example, the storage and retrieval of quantum states, a crucial requirement for emergent quantum technologies [107]. In this and other similar demonstrations, the low emitter-photon interaction probabilities necessitated the use of dense atomic ensembles, as was also the case when the atoms were weakly coupled to a waveguide [108–110]. Using waveguides to enhance the emitter-photon interaction can bring EIT to the single-photon and single-emitter level [106, 111], and such a system could form the basis for a single-photon all-optical switch [112].

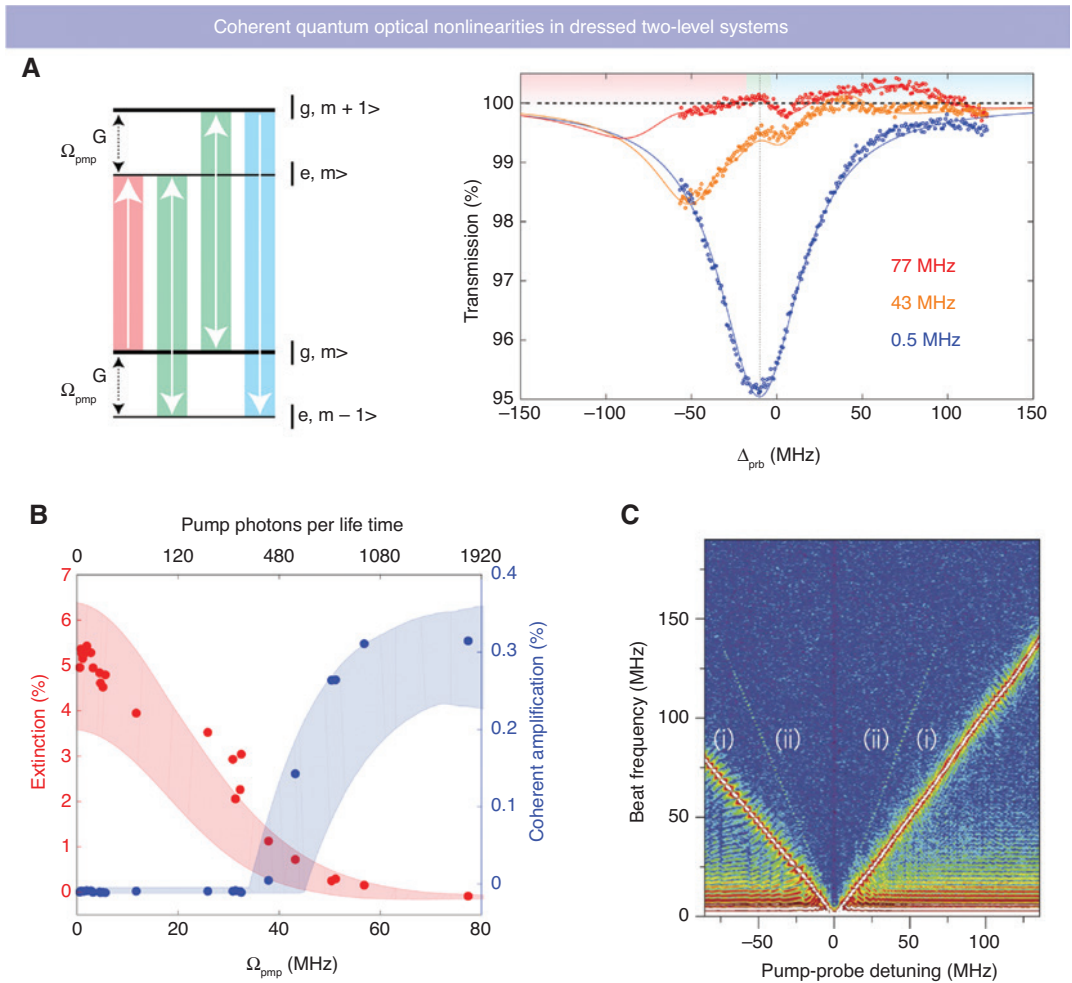

**Figure 5:** Coherent quantum optical nonlinearities in dressed two-level systems.
(A) The transmission spectra of photons scattering from a single organic molecule embedded in a nanoguide evolves nonlinearly as a function of the strength of a control beam with Rabi frequency $\Omega_{pmp}$ [23]. The features of the red and orange curves can be understood in terms of the three available transitions between the states of the system, as shown to the left; each manifold of states is described by the state of the TLS and the photon number, here $|e\rangle$ or $|g\rangle$ and $|m\rangle$, respectively, and the level splitting is given by the generalized Rabi frequency $\Omega_{pmp}^G = \sqrt{\Delta_{pmp}^2 + \Omega_{pmp}^2}$. Here, a Stark shift of the resonance (red), a coherent energy transfer between the signal and control photons (green), and a coherent amplification of the signal photons (blue) are observed. (B) The nonlinear dependence of the coherent extinction and amplification as a function of control beam strength. Adapted from [23]. (C) Four-wave mixing observed using a single organic molecule as the nonlinear medium. This nonlinear signal manifests at twice the beat frequency (ii) of the scattered signal (i). Adapted from [97].

Other coherent effects in three-level systems can be used to control the transport of photons. Population inversion of a single molecule, for example, allowed it to act as a quantum optical transistor [113, 114], coherently attenuating or amplifying a stream of photons. Interestingly, it is possible to form an optical transistor using a three-level Λ system without the need for population inversion. Rather, the coherent reflection and transmission outlined in Section 2.1 can be used, in conjunction with a gate pulse that effectively couples or decouples the emitter from the waveguide, bringing the transistor to the single-photon level if the emitter is efficiently coupled to the photonic mode [76, 111] (c.f. Figure 6B). Similarly, control over the state of a ladder-type emitter coupled to a waveguide can switch, or even impart a $\pi$ phase shift, to a guided photon [115]. Theories of the interaction of a few photons with three (or higher) level emitters predict the generation, or even engineering, of entangled photonic states. For example, two distinguishable photons can be entangled as they scatter from a ladder-type emitter, with the degree of entanglement depending on the spectral content of the photons [116]. The subsequent scattering of additional photons could thus be used to create large photonic entangled states, as shown in Figure 6C. The controlled rescattering of selected photon pairs from the entangled chain would create a photonic cluster state [105], which is a required resource in one-way quantum computating architectures [117].

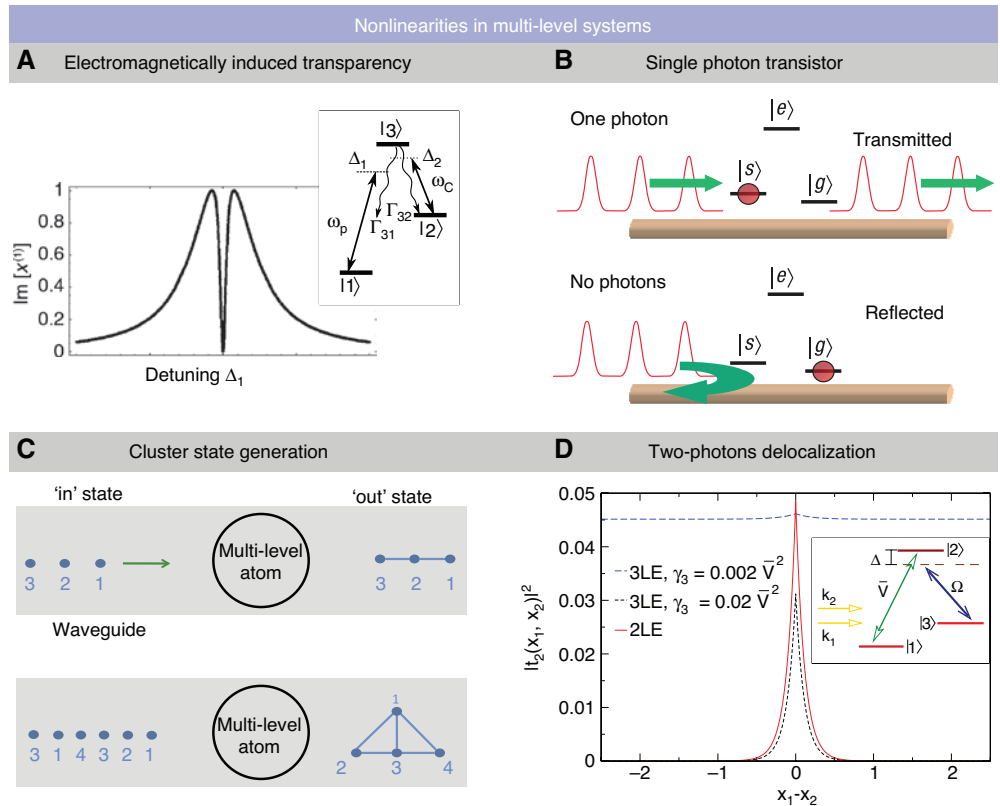

**Figure 6:** (A) Generic Λ-type scheme (left) and spectrum (right) for EIT where a probe field of frequency $\omega_p$ and control field of frequency $\omega_c$ interact with a quantum emitter. In the presence of the control beam transition $|1\rangle \to |3\rangle$ is inaccessible and hence the absorption spectrum sharply falls to 0 on resonance. Adapted from [104]. (B) Schematic diagram of a single-photon transistor based on a three-level emitter. The storage of a gate pulse containing zero or one photon conditionally spin-flips the state of the emitter, depending on the photon number. A subsequent incident signal field is then either transmitted or reflected depending on the state of the emitter. Adapted from [76]. (C) Example of photonic cluster state generation by sequential scattering from a multilevel emitter. In the top panel, three un-entangled photons sequentially scatter from an emitter, creating a three-photon matrix product state. In the bottom panel, four photons scatter from the emitter. After the fourth photon scatters, the first and third photons are re-scattered, creating a two-dimensional projected entangled pair state. Adapted from [105]. (D) Calculated two-photon spatial probability distribution after the scattering of two photons off a driven three-level emitter (energy-level scheme shown in inset). In the low-loss limit (i.e. $\gamma_3 \approx 0$), the two-photon state is delocalized. As $\gamma_3$ increases, the behavior of the driven system begins to resemble that of a two-level emitter. Adapted from [106].

Photonic bound states (c.f. Section 2.2) are also created when few-photon coherent states scatter from multilevel emitters [118]. In contrast to the scattering from two-level emitters, the presence of additional levels provides a route to the controlled shaping of the bound state. Driving a resonance of a Λ-type emitter, for example, can delocalize the two-photon wavepacket formed as the pair of photons scatter from the second transition (Figure 6D), paving a route toward control of the temporal properties of two-photon wavepackets [106]. Similarly, in an $N$-level emitter, multi-photon bound states can be made to destructively interfere with the standard multi-photon transmission, effectively suppressing multi-photon transmission and leading to a photonic blockade without the need for a cavity [88].

# 4 Nonlinear response of coupled multi-emitter systems

One of the most difficult and potentially most rewarding challenges in modern quantum optics is the scaling of individual elements into more complex quantum systems. This section covers recent works on multi-emitter waveguide quantum electrodynamics (QED), the vast majority of which are theoretical in nature, and which reflect both the difficulty and the reward of this undertaking.

Practically, the task is that the emitters of such a system simultaneously fulfill the following requirements: (i) They must all couple to the same guided mode, and in general this coupling should be efficient. (ii) The emitters should all interact coherently with passing photons (i.e. $\Gamma_{deph} \approx 0$, c.f. Section 2.1). (iii) The emitters should emit at a similar transition frequency and have similar linewidths. (iv) Ideally, the emitters could be individually addressed such that their relative coupling can be controlled (e.g. by local electrical gating).

As we outlined in Section 2, conditions (i) and (ii) have been met for a single emitter coupled to a waveguide, and the current experimental challenges lie in meeting criteria (iii) and (iv). First efforts in this direction involved the entanglement of two implanted Si-vacancy defects coupled to a single waveguide, first using a remote Raman control scheme [26] and then directly via strain tuning [119]. A signature of the entanglement between the two emitters was observed in their photon correlations. Similarly, superradiance has been observed using ensembles with a mean number of atoms ≤6 coupled to waveguide [33, 66], although here the emitters could not be individually addressed. In contrast, microelectrodes were shown to Stark-shift the resonance of many individual molecules coupled to a nanoguide [120, 121] (see Figure 7A). These experiments demonstrate that an efficient and controllable coupling of multiple solid-state emitters on a photonic chip is within reach [123].

Coupling quantum emitters with waveguides has been the focus of intense theoretical studies in recent years, resulting in prediction of both emergent many-body phenomena and new protocols for quantum information technology. This is already exemplified in two-emitter systems, which, even when coupled to a lossy waveguide, can result in sub- and super-radiant states and allow for two-qubit gates [124] and entanglement generation [125, 126]. Conversely, the high efficiency with which individual solid-state emitters can couple to a waveguide allows for the generation of Bell states [127], the entanglement and coupling of qubits that operate at vastly different frequencies, such as superconducting qubits with single molecules [128] or QDs [129]. Similarly, the input-output formalism was used to study quantum interference and photon statistics in a two-qubit system, demonstrating the complex dynamics that result from the quantum jumps of the emitters [130].

Concurrently, frameworks describing photonic transport through waveguides coupled to many emitters have been developed. Models focusing on specific aspects of this transport have quantified both the photon-photon [131] and emitter-emitter [132] entanglement that is generated as photons interact with the emitters, showing that this entanglement is more robust and efficient in chiral geometries. Interestingly, researchers have predicted that both chiral geometries [133] and stronger photonic correlations [134] lead to lower propagation losses, even if the emitters are weakly coupled to the waveguide or in the presence of disorder. Scattering from multiple emitters can also give rise to complex Fano shapes in the transmission and reflection spectra [135], providing a new route toward the control of photon statistics [79]. At the same time, efficient multiple scattering events between the emitters have been predicted to allow for normal mode-splitting without the need for cavities, and for the emergence of localized excitations [59] (c.f. Figure 7A) and "fermionic" subradiant modes that repel one another [136].

Many different effects and geometries of one-dimensional multi-emitter systems have been modeled. These include the addition of evanescent emitter-emitter coupling for closely spaced emitters [59, 137] and the inclusion of different decoherence mechanisms and inhomogeneous broadening [110]. Likewise, Pivovarov et al. developed a general microscopic model to describe single-photon scattering from a chain of multilevel emitters that describes both ordered and disordered geometries, and including both elastic and inelastic scattering channels [138]. Das et al., meanwhile, modeled the dynamics and amplitudes of the scattering of photons from multilevel emitters in the low excitation limit, working in the Heisenberg picture [139]. In this same low-excitation regime ($\langle \hat{\sigma}_{ee} \rangle = 0$), the Green's tensor approach outlined in Section 2 was generalized to $N$ emitters. For two-level emitters, in this Green's function formalism, the generalized equation of motion for the emitter coherences is

$$\dot{\hat{\sigma}}^i_{ge} = i\left(\Delta_A + i\frac{\Gamma'}{2}\right)\hat{\sigma}^i_{ge} + i\Omega^i + i\sum_j g_{ij} \hat{\sigma}^j_{ge}, \qquad (16)$$

where zero dephasing is also assumed. Here, the superscripts $i$ and $j$ refer to specific emitters, $\Gamma'$ is the rate of emission into the non-guided modes, and $g_{ij}$ is the dipole-projected Green's function as defined in Eq. (7). Note, however, that in this case $g_{ij}$ controls the interactions between the emitters, which can be either dispersive or dissipative, depending on whether the real or imaginary component of $\mathbf{G}(\mathbf{r}_i, \mathbf{r}_j)$ dominates. The general nature of this approach means that it can be applied to a variety of nanophotonic structures, including emitter chains in unstructured waveguides, cavities, and photonic crystals, as was done in [60] (Figure 7B). Protocols, based on these theories, have begun to emerge, including schemes for quantum computation [6] and the efficient generation of multi-photon states [140].

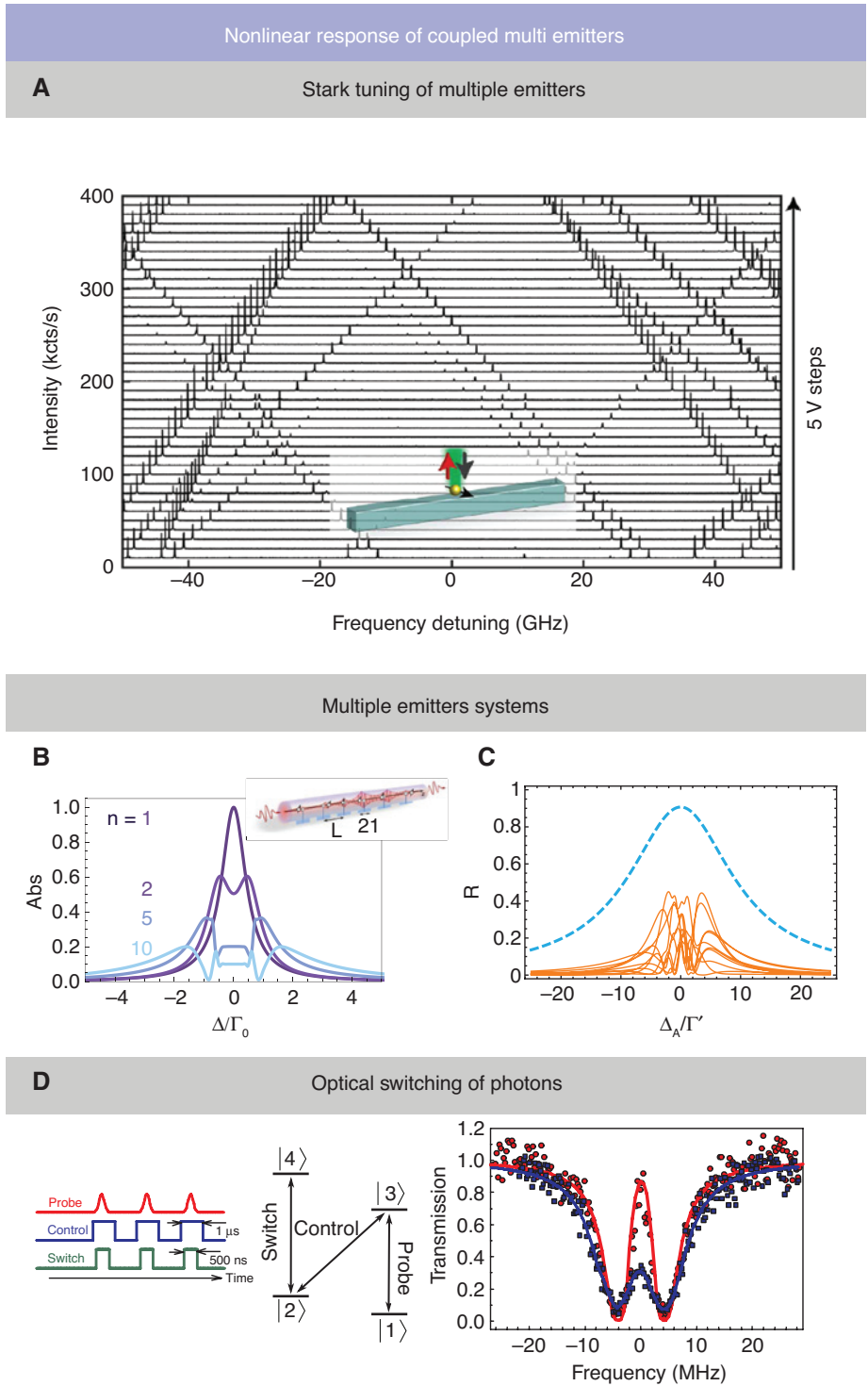

**Figure 7:** (A) Stark tuning of many organic molecules found within a single confocal excitation spot, as shown in the inset. Spectra taken at different Stark voltages demonstrate that it is possible to tune multiple emitters into resonance with one another on a photonic chip. Adapted from [120]. (B) Absorption spectrum calculated for a chain of well-coupled $N$ emitters ($\beta=0.99$) arranged at arbitrary positions about a regular lattice spacing $L = 2.75$ in a nanoguide (see inset). As the number of emitters increases (dark to light hues), normal-mode splitting is observed, while for $N=10$ narrow peaks near zero detuning $\Delta=0$ emerge because of the presence of subradiant modes. Adapted from [59]. (C) Reflection spectra for 20 atoms interacting through the guided modes of an unstructured waveguide. The dashed blue line represents a regular separation between the atoms of $\lambda/2$. The orange curves show 10 different spectra obtained by randomly placing the atoms along the nanostructure. Adapted from [60]. (D) Nonlinear optical switch based on four-level emitters coupled to a fiber. The timing sequence for the probe, control, and switch fields (right) and the corresponding level scheme of the emitters (center). Left: In the absence of the switch field, the probe field is largely transmitted through the fiber (red data). A strong suppression of the transmission is observed when the switch field is on (blue data). Adapted from [122].

The large nonlinearity inherent to quantum emitters manifests in novel fashion when many multilevel emitters are coupled via waveguides. Emitters in the EIT configuration [141], for example, can exhibit a giant Kerr nonlinearity [142]. Using a hollow-core photonic crystal fiber as a waveguide, to which they coupled a large ensemble of Rydberg atoms, researchers were able to exploit this Kerr nonlinearity for few-photon switching [122] (Figure 7C). A recent theoretical treatment of this system predicts that, in the ideal case, single-photon switching is possible, and studies the nonlinear evolution of two-photon wavepacket (c.f. Section 2.2) [143]. Photons traveling through such a multi-emitter system have also been predicted to crystalize, forming fermionic excitations that repel one another and providing an additional route to the creation of a pure single-photon source and enabling the study of complex quantum phase transitions [144]. Interestingly, the long-range interactions in such multi-emitter-waveguide systems are expected to give rise to nonlocal optical nonlinearities [145], providing yet another route to the creation of photonic bound states [146].

## 5 Conclusions and outlook

Advances in the growth and preparation of solid-state emitters and nanofabrication protocols have, in recent years, brought coherent light-matter interactions to quantum photonic chips. For single-emitter systems, these advances have allowed for the observation of a variety of nonlinear optical effects at, or near, the single-photon level, bringing a host of classical and quantum functionalities tantalizingly within reach. At the same time, increasingly complex theories have been developed that model the coupling of multiple emitters via long-range, waveguide-mediated interactions. Such multi-emitter systems have been predicted to support exotic new quantum phases of light and may enable efficient new quantum information technologies. Experimentally, we stand at the cusp of this exciting field, with multi-emitter photonic architectures just around the corner.

**Acknowledgments:** P.L., H.J., S.F.S, and N.R. gratefully acknowledge financial support from the Danish National Research Foundation (Center of Excellence Hy-Q) and the Europe Research Council (ERC Advanced Grant SCALE). S.G. and V.S. acknowledge financial support from the German Federal Ministry of Education and Research, co-funded by the European Commission (project RouTe), project number 13N14839 within the research program "Photonik Forschung Deutschland".